\documentclass[journal]{IEEEtran}
\ifCLASSINFOpdf
\else
\usepackage[dvips]{graphicx}
\fi
\usepackage{url}
\usepackage{graphicx, dblfloatfix}
\graphicspath{{./figures/}}
\usepackage{cite}
\usepackage{hyperref}
\hypersetup{
	colorlinks=true,
	linkcolor=blue,
	anchorcolor=black,
	citecolor=green,
	filecolor=magenta,      
	urlcolor=cyan,
	bookmarks=true,}
\usepackage{multirow}
\usepackage{flushend}
\usepackage{pifont}
\usepackage{mathtools, amsmath, amssymb, xfrac, tabularray}
\usepackage{color, tabularray}
\usepackage[linesnumbered,ruled,vlined]{algorithm2e}

\SetCommentSty{mycommfont}
\SetKwInput{KwInput}{Input}
\SetKwInput{KwOutput}{Output}

\begin{document}	
\title{Towards Efficient and Accurate CT Segmentation via Edge-Preserving Probabilistic Downsampling}
\author{Shahzad Ali, Yu Rim Lee, Soo Young Park, Won Young Tak, and Soon Ki Jung
\thanks{This work was supported by the Innovative Human Resource Development for Local Intellectualization program through the Institute of Information \& Communications Technology Planning \& Evaluation (IITP) grant funded by the Korean government (MSIT)(IITP-2024-RS-2022-00156389) and also supported by the Digital Innovation Hub project supervised by the Daegu Digital Innovation Promotion Agency (DIP) grant funded by the Korea government (MSIT and Daegu Metropolitan City) in 2024(No. DBSD1-04, Smart Management System for Preventing Deaths of Older People Living Alone Based on Automatic Meter Reading and CCTV Access Information). \textit{(Corresponding author: Soon Ki Jung.)}}
\thanks{Shahzad Ali and Soon Ki Jung are with the School of Computer Science and Engineering, Kyungpook National University, Daegu, South Korea (e-mail: shahzadali@knu.ac.kr; skjung@knu.ac.kr).}
\thanks{Yu Rim Lee, Soo Young Park, and Won Young Tak are with the Department of Internal Medicine, Kyungpook National University Hospital, College of Medicine, Kyungpook National University, Daegu, South Korea (e-mail: deblue00@naver.com; psyoung0419@gmail.com; wytak@knu.ac.kr).}}

\markboth{Journal of \LaTeX\ Class Files, Vol. xx, No. x, April 2024}
{Shell \MakeLowercase{\textit{Ali et al.}}: Bare Demo of IEEEtran.cls for IEEE Journals}
\maketitle

\begin{abstract}
Downsampling images and labels, often necessitated by limited resources or to expedite network training, leads to the loss of small objects and thin boundaries. This undermines the segmentation network's capacity to interpret images accurately and predict detailed labels, resulting in diminished performance compared to processing at original resolutions. This situation exemplifies the trade-off between efficiency and accuracy, with higher downsampling factors further impairing segmentation outcomes. Preserving information during downsampling is especially critical for medical image segmentation tasks. To tackle this challenge, we introduce a novel method named Edge-preserving Probabilistic Downsampling (EPD). It utilizes class uncertainty within a local window to produce soft labels, with the window size dictating the downsampling factor. This enables a network to produce quality predictions at low resolutions. Beyond preserving edge details more effectively than conventional nearest-neighbor downsampling, employing a similar algorithm for images, it surpasses bilinear interpolation in image downsampling, enhancing overall performance. Our method significantly improved Intersection over Union (IoU) to 2.85\%, 8.65\%, and 11.89\% when downsampling data to 1/2, 1/4, and 1/8, respectively, compared to conventional interpolation methods.
\end{abstract}


\IEEEpeerreviewmaketitle

\vspace{-7mm}
\section{Introduction}
\IEEEPARstart{S}{emantic} segmentation plays a pivotal role in medical image analysis by differentiating organs and anatomical structures by assigning a definitive class to each pixel, producing \textit{hard labels}. Despite recent advancements that benefit from large datasets and significant computational power, such dependencies pose challenges for researchers with constrained budgets. The necessity for full-resolution image processing demands considerable computational resources and memory, limiting broader participation. In response, lightweight networks with fewer trainable parameters have been proposed, facilitating operation on mid- to low-range devices at the expense of performance \cite{ma18ShuffleNet, mehta18ESPNetV1, howard19MobileNetV3, tan21EfficientNetV2, yu21BiSeNetV2, yu22MiniSeg, gao21MSCFNet, alcover23soft, xu23EffTrans}. As alternatives, downsampling or cropping datasets reduce resource demands but may compromise accuracy. Although cropping can be advantageous for images featuring symmetrical elements, repetitive patterns, or uniform class distributions, it may not suit all scenarios. The risk of excluding vital information when processing diverse datasets can impair the network's understanding of the global context, underscoring the importance of selecting an appropriate downsampling technique \cite{alcover23soft}. Nearest-neighbor interpolation, commonly used for labels, efficiently generates binary labels. Conversely, bilinear interpolation, preferred for image downsampling, can introduce aliasing, moiré patterns, or half-pixel offsets, which soften details and create jagged edges. Other methods, such as Lanczos or bicubic interpolation, may provide superior results but at a higher computational cost. 

Recently, there has been a growing interest in \textit{soft labels}, which marks a significant shift from traditional hard labels. Soft labels by assigning a probability distribution (i.e., continuous values between 0 and 1) across possible classes for each pixel, introduce uncertainty \cite{alcover23soft, lourencco21using, zhang23soft} and allow networks to capture better inherent ambiguity and fuzziness of edges and objects, potentially enhancing performance. Soft labels can be generated through various techniques, including label smoothing \cite{gros21softseg, li2020superpixel, muller19does, lukasik20a, lienen21label} and augmentation methods \cite{garcea23DataAug}. For example, cropping/resizing images using bilinear interpolation or introducing Gaussian noise can create uncertainty around object boundaries. Furthermore, soft labels can stem from annotators' disagreements regarding object boundaries in intra- and inter-rater annotations \cite{zhang23soft}. Averaging or fusing such annotations produces soft labels, while the majority voting makes hard labels. Empirically, loss functions tend to steer network predictions towards extreme values (0 or 1) rather than closely aligning with target soft labels, affecting class probability estimations, learning trajectories, and performance metrics. The success of a segmentation framework largely hinges on the labeling scheme, input resolution, network architecture, data augmentation, and loss function definition. This work introduces the novel Edge-preserving Probabilistic Downsampling (EPD) method, which embraces class uncertainty and variability to generate downsampled soft labels, enabling more efficient computational resource utilization and shortened training durations. Our contributions are as follows:
\begin{itemize}
	\vspace{-0.5mm}
	\item We propose a novel EPD method to produce reliable soft labels, thereby enhancing the performance of multi-class semantic segmentation on CT images.
	\item We adapt EPD for image downsampling to preserve edges and intricate details better.
	\item Our experimental findings, based on an in-house CT dataset, reveal consistent performance enhancements upon substituting conventional downsampling methods.
\end{itemize}

\begin{figure}[t]
	\centering
	\includegraphics[width=0.70\columnwidth]{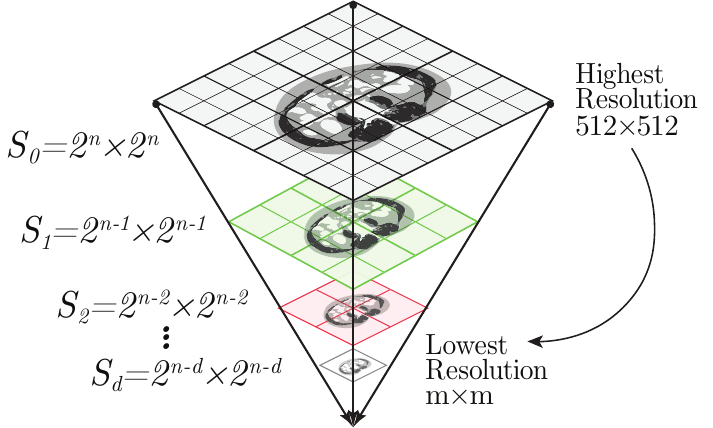}
	\vspace{-2mm}
	\caption{EPD exhibits an inverted pyramid-like structure with the highest and the lowest resolutions at the base and apex, denoted as $S_0$ and $S_{d}$, respectively. Here, $n=9$, each level halves the preceding resolution beginning from the original resolution of ${512\times512}$.}
	\label{fig01_pyramid}
	\vspace{-2mm}
\end{figure}

\section{Methodology}
\vspace{-2mm}
\subsection{Problem Definition}
Given a dataset ${\mathcal{D}={\{(x_i,y_i)\}}_{i=1}^{N}}$ containing $N$ pairs of data and corresponding labels. Each pair comprises a CT image slice, denoted as ${x\in \mathbb{R}^{H\times W}}$, and its corresponding label ${y\in \mathbb{W}^{H\times W}}$, where $H$ and $W$ represent height and width. Typically, $y$ is a hard label, with each pixel $y(i,j)$ representing a class index $c\in C$, with C being the total number of classes. $c=0$ represents the background, while ${c>0}$ represents foreground classes. By computing class probabilities, we propose calculating soft values to construct a downsampled \textit{soft label} from the original high-resolution hard label. This transformation from a hard to a soft label can be expressed as ${y_{h}\in \mathbb{W}^{H\times W} \rightarrow y_{s}\in \mathbb{R}^{H\times W\times C}}$, where $0\le y_{s}\le1$. We postulate the high-resolution label with size ${2^n\times2^n}$ at the base of an inverted pyramid as $S_{0}$ (Fig. \ref{fig01_pyramid}). Subsequently, we derive a downsampled label $S_{d}$ at the $d$-th level ${(1\leq d\le n)}$ from the base.

\vspace{-4mm}
\subsection{Downsampling Algorithm}
\subsubsection{For Labels} 
\begin{figure}[t]
	\centering
	\includegraphics[width=0.75\columnwidth]{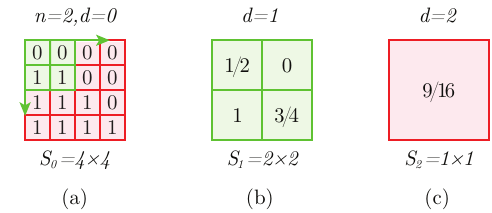}
	\vspace{-2mm}
	\caption{Proposed downsampling for an original-resolution label. (a) For simplicity, input label $S_0$ is binary, and the probability of foreground class $p_1$ is calculated. (b) $d=1$ downsamples the input resolution to half. (c)  $d=2$ downsamples the input resolution to quarter. The downsampling effect is prominent for windows containing edge pixels.}
	\label{fig02_example}
	\vspace{-4mm}
\end{figure}

Each pixel in downsampled label $S_{d}$ is derived from a window $w$ of pixels in the high-resolution label $S_{0}$. At level ${d=1}$, $S_1$ represents the downsampled label having half-resolution (${2^{n-1}\times2^{n-1}}$). Constructing $S_1$ involves a ${2\times2}$ window containing four $S_0$ pixels: ${S_{0}(2i-1,2j-1)}$, ${S_{0}(2i-1,2j)}$, ${S_{0}(2i, 2j-1)}$, ${S_{0}(2i, 2j)}$ (green window in Fig. \ref{fig02_example}(a)). The calculated class probabilities at the window level are associated with a single soft pixel, ensuring that the sum of all probabilities is always one, i.e., $\sum_{c}^{C}p_c=1$. This guarantees preserving the original class probability distribution in the downsampled labels. Notably, the absence of a class corresponds to $p_c=0$, a uniform region is represented by $p_c=1$, and a potential edge lies in $p_c\in(0,1)$. The effects of the proposed EPD method become evident in edge regions, producing an anti-aliasing effect. Fig. \ref{fig02_example}(b) presents a soft label $S_1$ with half-resolution calculated for an input binary label where the probability of foreground and background class is ${P_1=[\frac{1}{2},0,1,\frac{3}{4}]}$ and ${P_0=1-P_1=[\frac{1}{2},1,0,\frac{1}{4}]}$. For soft label of quarter-resolution $S_2$, the window size increases to ${4\times4}$ (red window in Fig. \ref{fig02_example}(a)), and probabilities are calculated similarly (Fig. \ref{fig02_example}(c)). Algorithm \ref{algo_downsample} presents the pseudocode for the proposed \textit{EPD} method.

\begin{algorithm}[b]
	\caption{The Proposed EPD for Labels}
	\smaller
	\label{algo_downsample}
	\DontPrintSemicolon
	\KwInput{High-resolution label $S_{0}$ of size ${2^n\times2^n}$, desired resolution ${2^{n-d}\times2^{n-d}}$}
	\KwOutput{Downsampled label $S_{d}$ of size ${2^{n-d}\times2^{n-d}}$}
	\BlankLine
	$h, \text{ }w \leftarrow 2^{n-d}, \text{ } 2^{n-d}$\;
	$S_{d}$ $\leftarrow$ 0\;
	\For{i $\leftarrow$ 0 \KwTo $h-1$}{
		\For{j $\leftarrow$ 0 \KwTo $w-1$}{
			\For{k $\leftarrow$ 0 \KwTo C}{
				\tcp{\smaller{get a window of pixels from $S_{0}$}}
				$w_{}(i,j,k) \leftarrow$ $S_{0}$ [$i\times d:(i+1)\times d, j \times d:(j+1)\times d,k$]\;
				\tcp{\smaller{update $S_{d+1}$ with class probabilities}}
				$S_{d}(i,j,k) \leftarrow {w_{}(i,j,k)}/{\sum w_{}(i,j,k)}$}
		}}
	\Return $S_{d}$
\end{algorithm}

\subsubsection{For Images}Extending the proposed method of class probability calculation to images is inefficient and necessitates further deliberation. Consequently, we propose an alternative approach, which entails substituting Step 6 in Algorithm \ref{algo_downsample} with the utilization of the average value of the window, denoted as ${S_{d}\leftarrow \frac{1}{|w_{}|}\sum_{}^{}w_{}(i,j)}$. EPD exhibits similarity (i.e., ${2\times2}$ window) with bilinear interpolation in constructing $S_1$. However, unlike bilinear interpolation, which utilizes weighted averages for computing a new value, we propose employing a simple average of all window pixels. This ensures computational efficiency and yields smoother handling of edge and uniform regions (Fig. \ref{fig03_downsampling}).

\vspace{-3mm}
\subsection{Training a Multi-class Segmentation Network}
A lightweight encoder-decoder network \cite[Fig. 2]{Ali2022} is employed in a fully supervised manner for multi-class segmentation. The encoder contains eleven \textit{Res-Conv} blocks and five \textit{MaxPool} layers. The decoder includes four \textit{Res-Conv} blocks along with five \textit{ConvTranspose} layers. The \textit{Res-Conv} block aggregates the inputs and extracted feature maps, with the learnable-weighted spatial and channel attentions, in an element-wise manner to produce the block output. The network outputs probabilities using a ${1\times1}$ convolutional layer followed by a \textit{softmax} layer. It undergoes training by minimizing the mean absolute error (MAE or $L_1$) and the dice similarity coefficient ($L_{dsc}$) losses defined as:
\begin{align}
	\label{eq_loss_l1}
	L_1 &= \frac{1}{N} \sum_{i}^{N}{{|y_{i}-\widehat{y}_{i}|}},\\[-0.75em]
	\label{eq_loss_dsc}
	L_{dsc} &= 1-\frac{2\sum_{i}^{N}{y_{i}.\widehat{y}_{i}}}{\sum_{i}^{N}{y_{i}}+\sum_{i}^{N}{\widehat{y}_{i}}},{\text{and}}\\[-0.1em]
	\label{eq_weight_loss_2}
	\mathbb{L}_{total}&=\sum_{L_{\tau}\in\mathbb{T}}\frac1{2\omega_\tau^2}{L}_\tau+ln{(1+\omega_\tau^2)},
\end{align}
where $y_{}$, $\widehat{y}_{}$, $N$ and $\mathbb{T}$ are the target soft labels, predicted soft labels, total number of labels, and a set of loss functions, respectively. 

In (\ref{eq_weight_loss_2}), the final loss $\mathbb{L}_{total}$ combines losses in $\mathbb{T}$ by assigning a weight $\omega$ to each loss term. This dynamic weighing is attributed to the task-dependent uncertainty varying across $\mathbb{T}$. This approach was initially proposed by \cite{Kendall2018} and further refined by \cite{Liebel2018} with positive regularization enforcement. Assuming each loss as a task results in an aggregated loss function $\mathbb{L}_{total}$ used for backpropagation in the network.

\begin{table*}
	\centering
	\caption{Performance Evaluation of the Proposed EPD Method.}
	\vspace{-2mm}
	\includegraphics{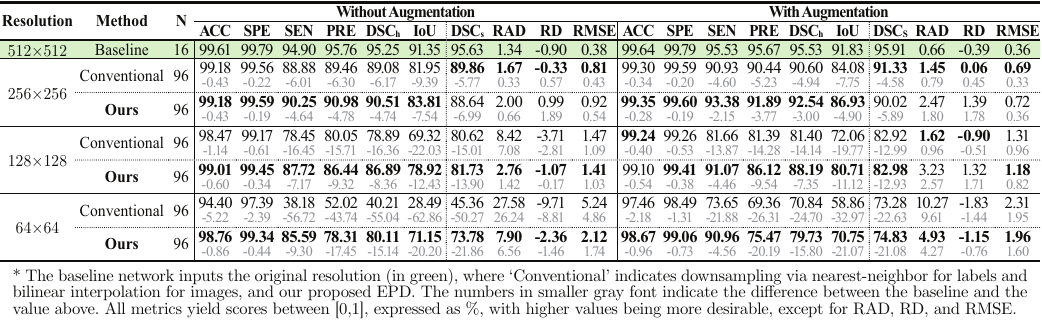}
	\label{table1}
	\vspace{-4mm}
\end{table*}

\vspace{-2mm}
\section{Experiments}
\vspace{-4mm}
\subsection{Dataset}
We trained and evaluated the proposed method on abdominal CT scans with a resolution of ${512\times512}$ from 57 patients at Kyungpook National University Hospital. These scans were retrospectively collected and labeled. 2D axial slices were extracted from six marked locations along the spinal column for each patient, resulting in 342 total slices. A radiologist subsequently conducted semi-manual labeling for the foreground classes: skeletal muscle (MUS), intermuscular adipose tissue (IMAT), saturated adipose tissue (SAT), visceral adipose tissue (VAT), and miscellaneous (MSC). The labeled set was randomly divided into $274$ training images ($80\%$) and $68$ validation images ($20\%$). The training set was augmented using horizontal flips, random cropping and resizing, and Gaussian blur.
\begin{figure}[t]
	\centering
	\includegraphics[width=0.9\columnwidth]{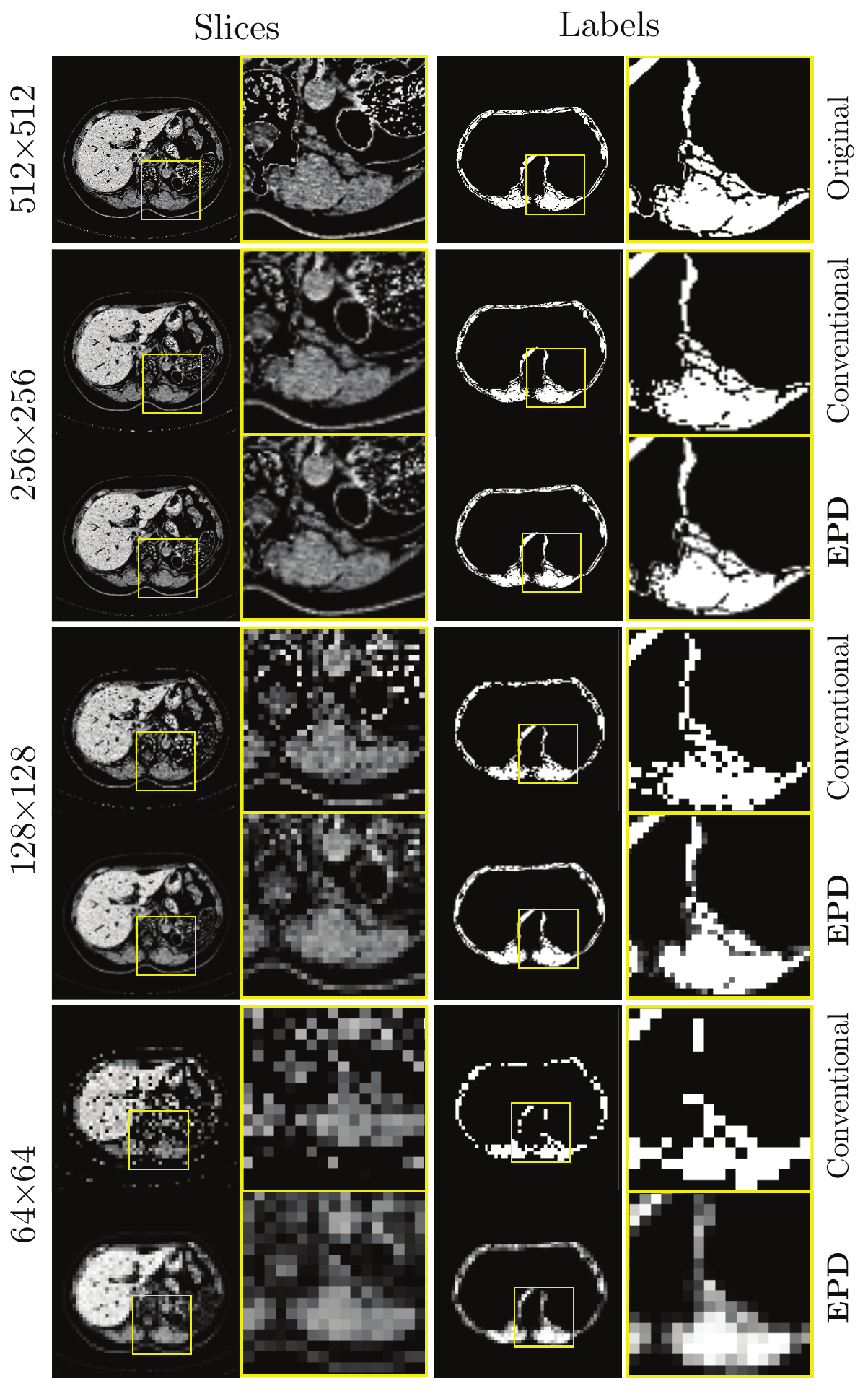}
	\vspace{-2mm}
	\caption{Application of EPD to the high-resolution inputs of $512\times512$. The conventional method employs bilinear and nearest-neighbor interpolation to downsample CT slices and labels, respectively. The window in yellow on the slices and labels is zoomed-in in the subsequent column.}
	\label{fig03_downsampling}
	\vspace{-4mm}
\end{figure}

\vspace{-4mm}
\subsection{Preprocessing}
Following \cite{Paris2020}, each CT slice is windowed to three different Hounsfield Unit (HU) ranges. These three channels, after downsampling, form a multi-channel input for the network. Similarly, the downsampled target labels are used for loss calculations. Fig. \ref{fig03_downsampling} illustrates the impact of downsampling on slices and labels. EPD transforms edges into floating-point probability values, which visually resemble anti-aliased edges. In contrast, nearest-neighbor interpolation disregards the presence of these pixels entirely, resulting in the loss of essential information, further leading to significant oversights in performance evaluation. When dealing with slices, averaging yields smoother edges while maintaining continuity better than bilinear interpolation.
	
\begin{figure*}[hb]
	\centering
	\includegraphics[width=2.0\columnwidth]{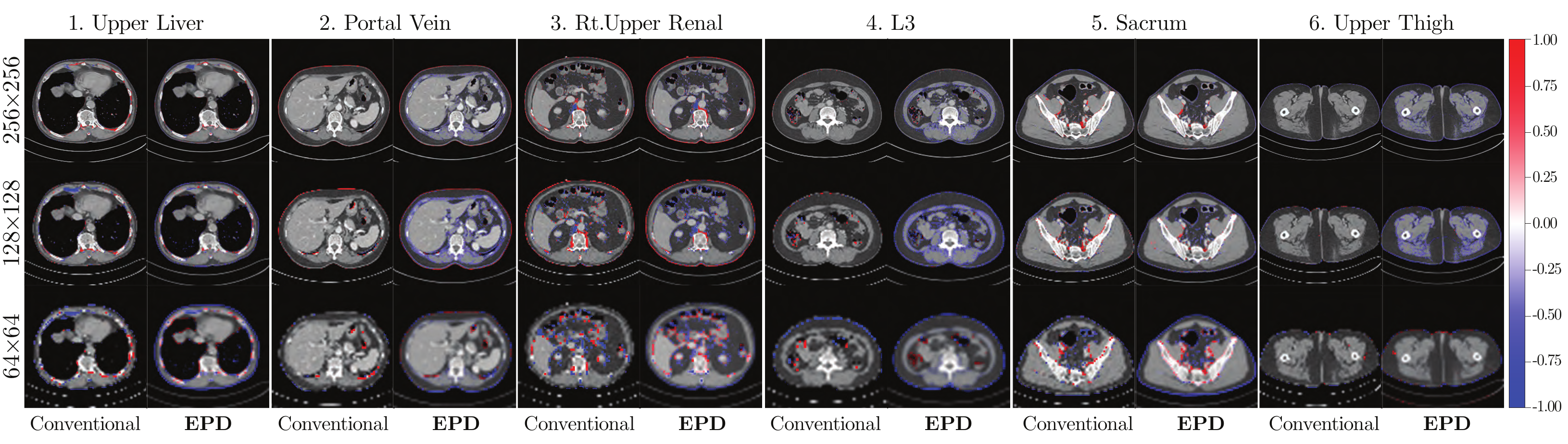}
	\vspace{-2mm}
	\caption{Probability discrepancy between the network prediction and target labels. These randomly selected CT slices belong to six marked locations in patients with varying genders and ages. The highest and lowest discrepancy is observed for the MSC and IMAT classes. Best viewed in color.}
	\label{fig04_results}
	\vspace{-4mm}
\end{figure*}

\vspace{-4mm}
\subsection{Evaluation Metrics}
We utilized two sets of metrics to gain a deeper understanding of the model's performance.

The first set encompassed accuracy (ACC), specificity (SPE), sensitivity (SEN), precision (PRE), dice similarity coefficient (DSC\textsubscript{h}), and intersection over union (IoU). The derivation of these metrics necessitates thresholding the predictions to identify the optimal threshold value over a range of values. We searched for the optimal threshold between [0,1], with a step size of 0.01, and selected the value producing the highest DSC\textsubscript{h}. The class label was assigned for soft targets based on the highest probability class or the first encountered class in case of a tie.

The second set of metrics comprised the soft dice similarity score (DSC\textsubscript{s}), relative absolute difference (RAD), relative difference (RD), and root mean squared error (RMSE). Unlike the hard metrics, the soft metrics do not necessarily work with one-hot encoded labels. RAD quantifies the proportion of change in prediction relative to the target, regardless of the sign. In semantic segmentation, distinguishing between under- and over-segmented predictions is crucial, so RD is also reported.

A critical factor in the evaluation process is handling missing class labels. Assigning zero values in such cases could underestimate the average metrics. Therefore, we opted to exclude these classes from the evaluation when applicable.
	
\vspace{-4mm}
\subsection{Implementation Details}
We trained the network from scratch on several resolutions for 100 epochs. The network depth and batch size were kept constant for all experiments to focus on downsampling effects. The only exception was the batch size for the baseline, which was reduced to 16 due to memory constraints. We utilized a fixed learning rate of $1e^{-3}$ and employed the AdamW optimizer. All experiments were conducted on four NVIDIA RTX A4000 GPUs, each with 16 GB of memory, using the PyTorch framework. No pre-training or transfer learning was employed.
	
\vspace{-4mm}
\section{Results and Discussion}
The performance of the proposed method is compared with that of conventional downsampling methods and a baseline model trained on the original resolution. For simplicity, bilinear interpolation for images and nearest-neighbor interpolation for labels are called \textit{conventional} downsampling methods. 

Table \ref{table1} details the results, both with and without data augmentations, each further showing the hard and soft metrics. The results from the baseline network serve as a reference for calculating the deviation, reported below each value in gray, of other methods' results. The conventional and EPD methods underperformed the baseline due to the inherent information loss during downsampling. However, EPD consistently outperformed the conventional method by significant margins, which increased with higher downsampling factors regardless of data augmentation. This trend is particularly evident for hard metrics in general. Performance gains of 2\% to 3\% were observed irrespective of the augmentation. It is noteworthy that while the proposed method showcases its potential even at the lowest resolution of 64$\times$64, augmenting the images did not yield discernible improvements at this scale. This might be attributed to the broader range of present probability values, potentially requiring more specialized and tailored augmentation techniques. 

Notably, EPD consistently resulted in DSC\textsubscript{h}$>$DSC\textsubscript{s}, indicating predictions with higher confidence. Conversely, the DSC\textsubscript{s} accounts for the entire probability distribution and penalizes deviations from target labels. Consequently, DSC\textsubscript{s} consistently remained higher for the low RMSE values. Furthermore, the best-performing scenarios often coincided with negative RD values, indicating under-segmentation tendencies.

Fig. \ref{fig04_results} illustrates the disparity between predicted and target label probabilities for slices at six spinal column positions. Red and blue denote higher and lower predicted values, respectively. Particularly noticeable at the junction of background and foreground classes, high disparity is observed around bones (MSC class), while low disparities are found in the MUS and IMAT classes. EPD consistently yields lower predicted probabilities for the IMAT class, reflecting challenges in discerning thin structures. This aligns with observations in Fig. \ref{fig03_downsampling}, where nearest-neighbor interpolation struggles with low-compact structures like IMAT class. Reflecting this in the final results poses challenges, as discerning whether information loss stems from inferior downsampling methods for target labels is challenging.

\vspace{-4mm}
\section{Conclusions}
The proposed EPD method demonstrates promising results in generating detailed soft labels, which hold the potential to enhance the learning and predictive capabilities of segmentation networks while reducing training time in resource-constrained environments. Seamlessly integrating this approach into various segmentation networks could unlock these benefits, fostering advancements in medical image analysis and beyond.

\bibliography{references.bib}
\bibliographystyle{IEEEbib}
\end{document}